\documentclass{aa}
\usepackage{txfonts}
\usepackage{graphicx}
\usepackage{booktabs}
\usepackage{amsmath}
\usepackage{colortbl}
\usepackage{gensymb}
\usepackage{subfig}
\usepackage{rotating}
\usepackage{amssymb}
\usepackage{latexsym}
\usepackage{ifthen}
\usepackage{enumitem}

\newcommand{\co}{CO(3-2) }

\def\lsimeq{\hbox{\raise0.5ex\hbox{$<\lower1.06ex\hbox{$\kern-1.07em{\sim}$}$}}} 
\def\gsimeq{\hbox{\raise0.5ex\hbox{$>\lower1.06ex\hbox{$\kern-1.07em{\sim}$}$}}} 

\begin{document}

   \title{Evidence for feedback in action from the molecular gas content in the $z\sim1.6$ outflowing QSO XID2028\thanks{Based on observations with the Plateau de Bure millimetre interferometer, operated by the Institute for Radio Astronomy in the Millimetre Range (IRAM), which is funded by a partnership of INSU/CNRS (France), MPG (Germany) and IGN (Spain).}}


   \author{M. Brusa
          \inst{1,2,3}\thanks{email:marcella.brusa3@unibo.it}
          \and
          C. Feruglio \inst{4,5,6}  
          \and
          G. Cresci \inst{7}
	  \and
          V. Mainieri \inst{8}
	  \and
          M. T. Sargent \inst{9} 
	  \and
         M. Perna \inst{1,3} 
       \and
        P. Santini \inst{6} 
       \and
       F. Vito \inst{1,3}
	  \and
   A. Marconi \inst{10} 
  \and
  A. Merloni \inst{2} 
  \and
   D. Lutz \inst{2}  
\and
  E. Piconcelli \inst{6} 
  \and
     G. Lanzuisi \inst{1,3}   
  \and         
   R. Maiolino \inst{11}  
 \and
  D. Rosario \inst{2}
\and
 E. Daddi \inst{12}
\and
     A. Bongiorno \inst{6}
	  \and
     F. Fiore \inst{6} 
    \and 
     E. Lusso \inst{7} 
          }

   \institute{Dipartimento di Fisica e Astronomia, Universit\`a di Bologna,
  viale Berti Pichat 6/2, 40127 Bologna, Italy 
 \and Max Planck Institut f\"ur Extraterrestrische Physik, Giessenbachstrasse 1, 85748 Garching bei M\"unchen, Germany 
\and INAF - Osservatorio Astronomico di Bologna, via Ranzani 1, 40127 Bologna, Italy 
\and IRAM - Institut de RadioAstronomie Millim\'etrique, 300 rue de la Piscine, 38406 Saint Martin d'H\`eres, France,
\and Scuola Normale Superiore, Piazza dei Cavalieri 7, I-56126 Pisa, Italy
\and INAF - Osservatorio Astronomico di Roma, via Frascati 33,   00044 Monte Porzio Catone (RM) Italy 
\and INAF - Osservatorio Astronomico di Arcetri, Largo Enrico Fermi 5, 50125 Firenze, Italy 
\and  European Southern Observatory, Karl-Schwarzschild-str. 2,  85748 Garching bei M\"unchen, Germany 
\and Astronomy Centre, Department of Physics and Astronomy, University of Sussex, Brighton, BN1 9QH, UK
\and Dipartimento di Astronomia e Scienza dello Spazio, Universit\`a degli Studi di Firenze, Largo E. Fermi 2, 50125 Firenze, Italy 
\and Cavendish Laboratory, University of Cambridge, 19 J. J. Thomson Ave., Cambridge CB3 0HE, UK
\and Laboratoire AIM, CEA/DSM-CNRS-Universit\'e Paris Diderot, IRFU/Service d’Astrophysique, B\^at.709, CEA-Saclay, 91191 Gif-sur-Yvette Cedex, France
     }

   \date{Received 9 December 2014; accepted 3 March 2015}

  \abstract

\abstract
{}
{Gas outflows are believed to play a pivotal role in shaping galaxies, as they regulate  both star
formation and black hole growth.  
Despite their ubiquitous presence, the origin and the acceleration mechanism of such powerful and extended winds  is not yet understood. 
Direct observations of the cold gas component  in objects with detected outflows at other wavelengths are needed to assess the impact of the  outflow on the host galaxy interstellar medium (ISM). }
{We observed with the Plateau de Bure Interferometer an obscured quasar at z$\sim$1.5, XID2028, for which the presence of an ionised outflow has been unambiguously signalled by  NIR spectroscopy. 
The detection of $^{12}$CO(3--2) emission in this source allows us to infer the molecular gas content and compare it to the ISM mass derived from the dust emission. We then analyze the results in the context of recent insights on scaling relations, which describe the gas content of the overall population of star-forming galaxies at a similar redshifts. }
{
The  Star formation efficiency ($\sim100$) and gas mass (M$_{\rm gas}=2.1-9.5\times10^{10}$ M$_\odot$) inferred from the CO(3-2) line depend on the underlying assumptions on the excitation of the transition and the CO-to-H2 conversion factor. However, the combination of this information and the ISM mass estimated from the dust mass suggests that the ISM/gas content of XID2028 is significantly lower than expected for its observed M$_\star$, sSFR and redshift, based on the most up-to-date calibrations (with gas fraction $<$20\% and depletion time scale $<$340 Myr).}
{Overall, the constraints we obtain from the far infrared and millimeter data suggest that we are observing QSO feedback able to remove the gas from the host.}

\titlerunning{Feedback in action in XID2028}
   \maketitle

%

\section{Introduction}

There are both theoretical (e.g. \citealt{Hopkins2008}) and observational (e.g. \citealt{Sanders1988,Yan2010}) arguments that 
support the notion that luminous star-forming galaxies (hereafter: `Starbursts') 
and luminous, unobscured Active Galactic Nuclei (AGN; hereafter luminous AGN or `QSO') are basically the same systems caught in different stages of the co-eval growth of (massive) galaxies and the Super Massive Black Holes (SMBH) sitting in their centres. 
In particular, Starbursts should trace objects caught in the rapid SMBH growth phase characterized by efficient Star Formation (SF), in a dust-enshrouded, dense environment, while the unobscured QSOs are systems radiating at the Eddington limit, where the SMBH is almost  fully assembled. 

Given that both SF and AGN activity are thought to be sustained by the availability of cold gas in galaxies
(see e.g \citealt{Menci2008,Vito2014_gas}), millimeter observations of molecular transitions are needed to directly probe the presence and state of this gas.
In the past decade, observations of  cold molecular gas reservoirs at high redshift (see  \citealt{CW2013} for a comprehensive review) turned out to be crucial in studying the gas content and consumption rate in both normal and peculiar systems.
For example, the gas properties of ``normal" galaxies are being investigated in increasing details up to high-z \citep{Tacconi2013,Genzel2014_gas,Sargent2014}, and as a function of many of the structural and physical properties of the systems (e.g. Star Formation Rate, SFR; stellar mass; colors; see e.g. \citealt{Genzel2014_gas,Sargent2014}). This has become possible thanks to the large investment of time at millimeter arrays, mainly the Plateau de Bure Interferometer (PdBI). 
In particular, it has been reported that, among massive systems,  (M$_{\star}>10^{10}$ M$_\odot$), the gas fraction increases across the main sequence (MS; defined between the SFR and the stellar mass of galaxies) at fixed redshift (see \citealt{Magdis2012a,Magdis2012b,Saintonge2012,Tacconi2013,Sargent2014}) and is hence  closely related to the Specific Star Formation rate (sSFR). This fits in a scenario where the redshift evolution of the sSFR is consistent with being driven by the gas fraction (see also \citealt{Lilly2013}).
Similar conclusions are reached in works involving dust fitting methods to derive the gas mass (see e.g. \citealt{Santini2014}).

The first molecular studies on local Ultra Luminous Infrared Galaxies (ULIRGs) and Submillimeter Galaxies (SMG) at higher redshifts, i.e. targetting objects in the `Starburst' phase, 
showed that these systems typically have a low molecular gas content with respect to their current SFR, or alternatively higher star-formation efficiencies.
Indeed, defining empirically the Star Formation Efficiency (SFE) as the ratio of the IR luminosity to the CO luminosity (in units of L$_{\rm \odot}$/(K km s$^{-1}$ pc$^2$)), `Starbursts' have SFE$>$200 (see e.g. \citealt{Daddi2010,Genzel2010}) larger than those observed  in normal star forming galaxies with the same molecular gas content  (\citealt{Tacconi2010}, SFE$\sim50-200$).
 In other words, their consumption time scale is shorter with respect to normal galaxies and they will exhaust their gas reservoirs in a short timescale ($\lsimeq100$ Myr). This is consistent with the hypothesis that `Starbursts' in general (and ULIRGs/SMGs in particular)  are objects at the peak of their SF activity in the heavily obscured phase.

On the other hand, high values of the  L$_{\rm IR}$/L'(CO) ratio have been also observed in high-z unobscured QSO host galaxies (SFE$>200$; e.g. \citealt{Solomon2005,Riechers2011,Riechers2011_QSOz3}), although, being a subsequent phase of `Starbursts' in the evolutionary sequence, their SFR is expected to be already substantially suppressed. 
In this case a significant fraction of the gas could have been previously removed during the `blow-out' phase, and the observed high SFE in unobscured QSOs can be  ascribed to region of residual, on-going SF, pointing towards a possible effect of 'positive feedback' on the galaxy from the AGN \citep{Silk2013,Zubovas2014}.  

   \begin{figure*}[!t]
   \centering
 \includegraphics[width=8.7cm,angle=0]{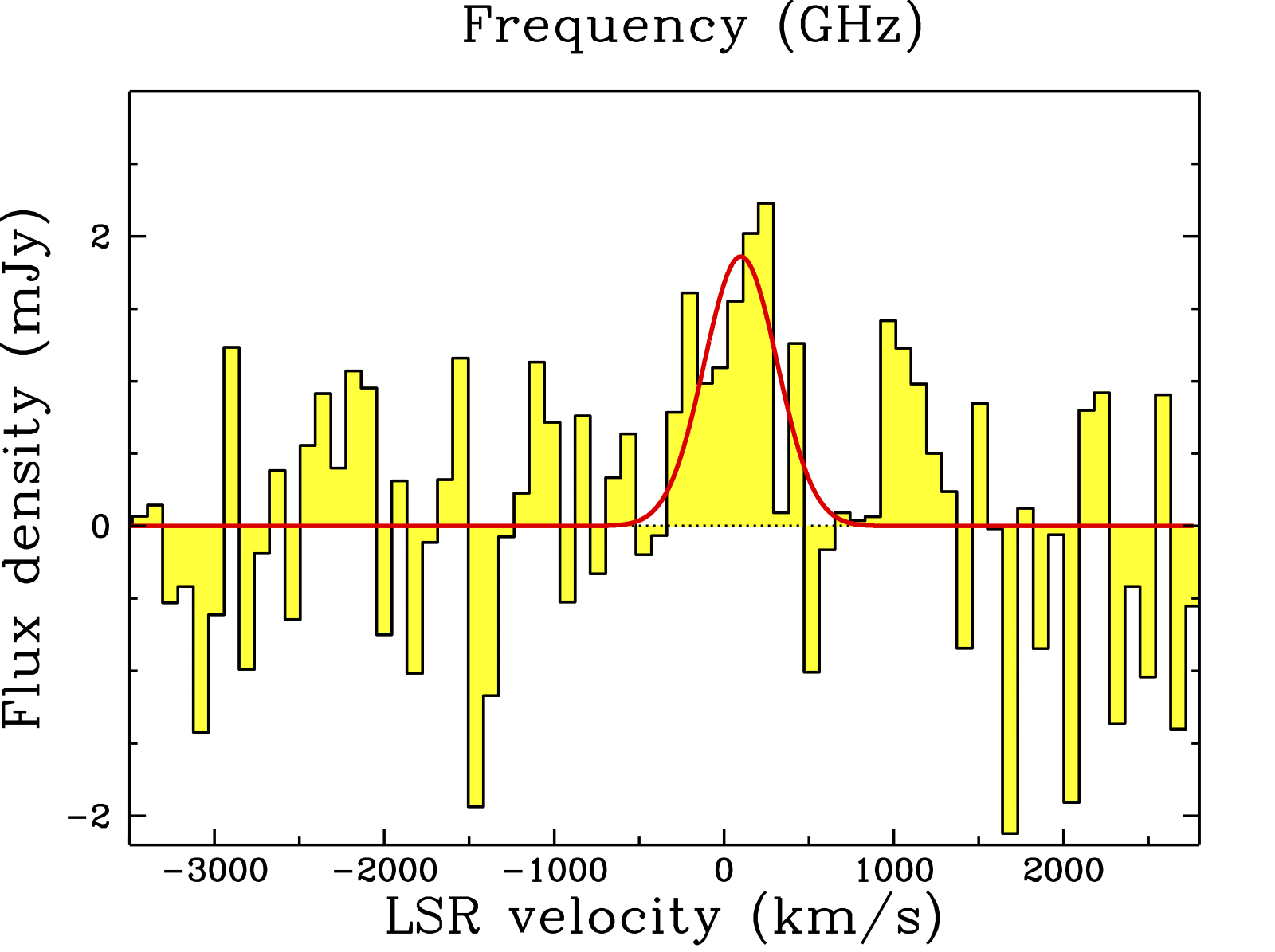}
\includegraphics[width=6.8cm]{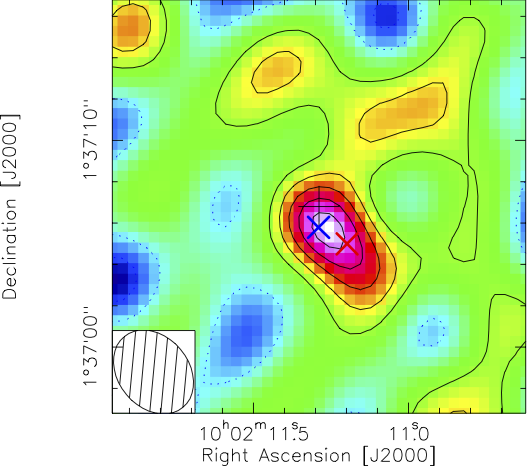}
      \caption{{\it Left panel}: spectrum of XID2028 integrated over the beam. The solid line shows a Gaussian fit with FWZI=770 km s$^{-1}$ and centered at the frequency corresponding to the redshift of the source. {\it Right panel}: integrated map of CO(3–2), in the channels corresponding to the ``systemic" peak of the line. Contour levels are 1$\sigma$ each ($\sigma$=0.23 Jy km s$^{-1}$). The synthesised beam is shown in the bottom-left corner. The black cross marks the phase center (i.e. the ACS nucleus). The blue and red cross mark the positions of the blue and red line components, as derived from our spectroastrometric analysis.
              }
         \label{coline}
   \end{figure*}
%

What is still missing for a full understanding of the results of  the aforementioned studies, in terms of the role of the  physical processes which govern the co-eval BH-galaxy growth, is a full characterization of the gas properties 
of objects  
caught in the short-lived ``transition'' phase between the Starburst and QSO stages.
This phase is expected to be characterized by gas reservoirs not yet depleted and by complex kinematics, including strong winds and outflows. 
\citet{Brusa2010}   proposed that sources in the `blow-out' phase at z$\sim1.5$ can be isolated on the basis  of their observed X-ray-to-optical-to-NIR colors and  presented the source XID2028 (z=1.5927), detected in the XMM-COSMOS survey, as the prototype of this class.
XID2028 is a luminous (L$_{\rm bol}\sim2\times10^{46}$ erg s$^{-1}$), mildly obscured QSO hosted in a massive galaxy, with M$_{*}\sim4.5\times10^{11}$ M$_\odot$ and a SFR$\sim270$ M$_\odot$ yr$^{-1}$ as measured by {\it Herschel} from PEP and SPIRE data \citep{Lutz2011,Bethermin2012}. 
At its center, XID2028 has a supermassive black hole with mass M$_{\rm BH}\sim3\times10^9$ M$_\odot$ \citep{Bongiorno2014}, which is accreting at
$\sim5$\% of its Eddington luminosity. 

The presence of a massive outflow in the ionized gas component of XID2028, traced by the [O III]$\lambda$5007 emission, 
has been unambiguosly and independently confirmed by X-shooter slit spectroscopy \citep{Brusa2015,Perna2015} and SINFONI J-band IFU observations:
in fact, XID2028  hosts one of the most massive ($\dot{M}_{ion}>250$ M$_\odot$ yr$^{-1}$, with v$>1500$ km s$^{-1}$) and most extended (out to scales of $\sim13$ kpc) outflows detected in a high-z QSO \citep{Cresci2015}. 
Most importantly, the outflow lies exactly in the center of a cavity in star forming regions in the host galaxy (as traced by narrow H$\alpha$ emission line map and rest frame U band imaging; see \citealt{Cresci2015}) thus suggesting that the wind is removing the gas from the host galaxy (`negative feedback’),
and at the same time is also triggering star formation by outflow induced pressure at the edges (`positive feedback’; e.g. \citealt{Zubovas2014}). XID2028  therefore represents   a test case to study QSO `feedback in action'.
However, the evidence of feedback  in this source mostly comes from measurements of the on-going star formation in the source traced by the narrow H$\alpha$ emission line that in principle may be affected by e.g. differential extinction effects in the host galaxy. 

Direct observations of the cold gas component in this galaxy are needed to assess 
whether the ionized outflow has an impact on the cold gas reservoir.
With this aim, here we present observations of the \co  transition of XID2028, redshifted to 2mm, obtained with the  PdBI Interferometer. 
We compare the gas masses derived from CO with that inferred from the dust mass and based on Far Infrared (FIR) data. 
These two methods allow us to investigate  whether AGN feedback has  already been effective in diminishing the cold gas mass in the host, or whether the feedback  phase is still associated with cold-gas-rich galaxies similarly to MS star-forming galaxies, with important consequences for galaxy-AGN coevolutionary models. 

The paper is organised as follows: Section 2 presents the PdBI observations and data analysis, Section 3 discusses the results, while Section 4 summarizes our conclusions. 
Throughout the paper, we adopt the cosmological parameters $H_0=70$ km s$^{-1}$ Mpc$^{-1}$, $\Omega_m$=0.3 and $\Omega_{\Lambda}$=0.7 (Spergel 2003). In quoting magnitudes, the AB system will be used, unless otherwise stated.  We adopt a Chabrier Initial Mass Function to derive stellar masses and SFRs for the target and comparison samples. The physical scale is 1"$\sim8.5$ kpc at the redshift of the source.

\section{Millimeter observations}

\subsection{Data reduction}
XID2028 was observed  with receivers tuned to a frequency of 133.37 GHz, corresponding to the expected frequency of the CO(3-2) emission line, with the PdBI
array in the most compact (D)  configuration. The observations were split in
3 tracks (31-May, 1, 6, June 2014).  The
system temperature (T$_{sys}$) was between 100 and 300 K, and water vapor  4-6
mm.  The quasar 1005+058 (0.3 Jy at 133.7 GHz) was used as a phase and
amplitude calibrator.  MCW349 (with a flux of 1.8 Jy) was used for absolute flux calibration,
which yields an absolute flux accuracy of about 5\% at the observed frequency. 
Calibration and mapping were done in the GILDAS environment.  The flagging of the phase visibilities was
fixed at $<45\degree$ rms. 

The total observing time was 5.6 hrs (3.06 hrs on source), 
for a total of 3673 visibilities available, before applying any flag.
We then removed one scan (3994 in 01-June track) due to problems with the tracking.
Data from antenna 1  from the 06-June track were not used in the final dataset due to the presence of a tuning parasite that produced a spurious signal at a frequency (133.21 GHz) close to the observed frame frequency of the CO(3-2) transition. After flagging of bad visibilities, the total on source time is 2.54 hours (six-antenna equivalent), and the 1$\sigma$ sensitivity is 1.36  mJy/beam in 20 MHz channels, for a total of 3052 visibilities.
The clean beam of the observations is 4.5"x3.4", with an angle of 38 degrees. 
The phase center of the data set was set to the HST position of the QSO nucleus  (RA=10:02:11.29, DEC=+01:37:06.79). 

\subsection{Analysis}
We estimated the 2 mm continuum by collapsing the line-free channels of the data set and fitting the visibilities.  
The continuum is not detected with a 3$\sigma$ upper limit on its flux of 0.3 mJy. 

The redshift of the host galaxy (z=1.5927) was adopted to convert the frequency to velocity space. 
Figure~\ref{coline} shows the line spectrum integrated over the beam.  
The line displays two peaks: one centered around the systemic redshift (FWHM$\sim$$550\pm200$ km s$^{-1}$ from a Gaussian fit), and another centered at $\sim1000$ km s$^{-1}$ (henceforth referred to as ``red feature"). The peak at the systemic position is significant at $5\sigma$, while the ``red feature" is at a lower significance ($\sim$3$\sigma$). 
Moreover, the ``red feature" peaks at $\sim133.0$ GHz,  close to a known parasite signal at 132.9 coming from antenna 4 and identified in all tracks. 
We created a new table flagging data on Antenna 4. The total exposure time decreased to 1.8hr  and the total number of visibilities to be used for the scientific analysis also considerably decreased. The red feature is not significant anymore (S/N$<3$). However, the significance of the detection over the systemic line also decreased at S/N$\sim$4. For this reason we decided to keep the full datasets in the analysis and, in the absence of deeper, more highly resolving observations which could confirm the presence of a second dynamically distinct component, we will consider the red feature as spurious.

The zero spacing flux estimated by fitting the averaged visibilities in the velocities range from -340 to +440 km s$^{-1}$ with a point source function is {\it S'(CO)}=1.6$\pm$0.3 mJy (5.3$\sigma$), and returns a centroid at (RA,DEC=10:02:11.24 01:37:05.48). 
The integrated flux over the full velocity range of the systemic line (with Full Width Zero Intensity, FWZI$\sim770$ km s$^{-1}$) is  therefore $\int S’(CO)dv$=
1.23$\pm$0.23 Jy km s$^{-1}$.
This measure depends only on the data calibration (including flagging of the Antennas) and does not depend on any other assumption, like e.g. masking, extraction region, ad-hoc centroid.
The quoted errors take into account the statistical errors of the {\it uv} plane fit and the errors on the absolute flux calibration (5\%). 
The right panel of Figure~\ref{coline} shows the integrated map over the 
systemic line emission.  
We verified that the flux extracted from the integrated map (S=1.55$\pm0.3$ mJy) on a region slightly larger than the beam, is in agreement with the one estimated by fitting the visibilities.

\begin{figure}[!t]
   \centering
 \includegraphics[width=8cm]{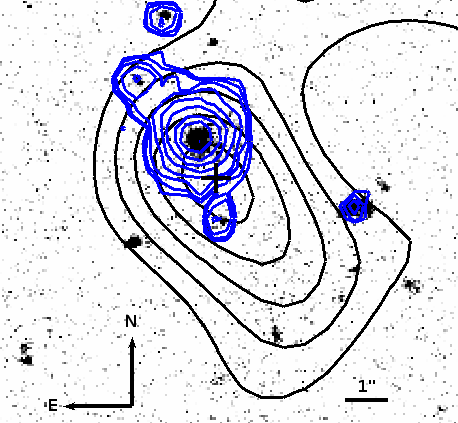}
      \caption{HST/ACS image (F814W filter) with superimposed K-band contours from CFHT (blue, arbitrary levels chosen to trace the whole K-band emission). Black contours represent CO(3-2) emission from the integrated map in the channels corresponding to the ``systemic'' peak of the line (same levels as in right panel of Figure 1; starting from 1 sigma).  The black cross (with associated ellipse) marks the line centroid. The image is about 10" across. The beam size is 4.5"x3.4", with an angle of 38 degrees.
              }
         \label{FigHST}
   \end{figure}

 Figure~\ref{FigHST} shows the HST/ACS image (background) with superimposed the contours  from the K-band image (blue), which should trace  the extension of the host galaxy. The black contours are from the map obtained on the line detected at the systemic position (e.g. from Figure 1 right, in steps of S/N, starting from 1$\sigma$) and the black cross marks the line centroid.

From both Figures 1 and 2 it is clear that the line peak is offset by $\sim1\arcsec$ from the QSO nucleus position. 
From previous observations  with the same phase calibrator (1005+058), we can exclude errors in the absolute astrometry. 
The error associated with the beam and the S/N of the source translates into a positional uncertainty of 0.46" x 0.36".  
We note, however, that the displacement may  be due to the limited {\it uv} coverage of the data, and that a CO-offset is typical of low S/N data (see e.g. \citealt{Casey2011}). Better signal-to-noise ratio and {\it uv} coverage are needed  to refine the location of the gas reservoir.

A dynamical mass can be estimated from the CO line width assuming a size ($\rm R$) and an inclination ($i$) of a rotating molecular gas disc.
The size can be inferred using the spectroastrometric technique \citep[and references therein]{Gnerucci2011,Carniani2013}, applied to the CO data cube. By integrating the CO data in the red (0,+400 km s$^{-1}$) and blue (-400,0 km s$^{-1}$) line channels, we measure a difference in the line centroids of $\sim1.5\pm0.2\arcsec$ (with an error of 0.3 pixels for each detection). The centroids of these detections are also shown in Figure 1 (right panel) as blue and red crosses to mark the blue and red line channels, respectively.
The measured shift corresponds to $\sim$13 kpc at the source redshift and translates to R$\sim$$6.5\pm0.8$ kpc, in agreement with the extension seen in the K-band data (see also \citealt{Cresci2015}).
Applying Equation 5 of \citet{Gnerucci2011}, we infer a M$_{\rm dyn}$(sin i)$^2$=4.5$\times10^{11}$ M$_\odot$ and, assuming an inclination of 60 deg, a M$_{\rm dyn}$$\sim$6.0$\pm2.3\times10^{11}$ M$_\odot$ once all the uncertainties in the quantities are taken in account. We will discuss in the following how this compares with the total mass derived from M$_{\star}$+M$_{\rm gas}$.

%

\section{Results and discussion} 

Deriving the luminosity L'CO[1-0] of the ground-state transition (which is generally regarded as the best indicator of the total gas reservoir) requires an assumption on r$_{31}$, the luminosity ratio between
the \co and the CO(1-0) transitions, which depends on the nature of the systems: a ratio of $\sim0.7-1$ is typically reported for SMG galaxies and QSOs (see \citealt{CW2013} and references therein), 
while  an average ratio of $\sim0.42$ has been determined for MS star-forming galaxies at a similar redshift as XID2028 (e.g. \citealt{Daddi2015}).

The \co luminosity of XID028 is  L'(CO[3-2])=1.9$\pm0.4\times10^{10}$ K km s$^{-1}$ pc$^2$ ($\sim9.3\pm2\times10^5$ L$_\odot$), following \citet{Solomon2005}. This value is in between the average values observed in this molecular transition for U/LIRGs (L[CO(3-2)]=$2.6\pm0.5$$\times10^{9}$ K km s$^{-1}$ pc$^2$) and SMGs (L[CO(3-2)]$4.4\pm1.1$$\times10^{10}$ K km s$^{-1}$ pc$^2$), as reported in the work of \citet{Iono2009}. On the other hand, the SFR ($\sim270$ M$_\odot$ yr$^{-1}$) and M$_\star$ ($\sim4.5\times10^{11}$ M$_\odot$) of XID2028 are consistent with those observed in a MS galaxy at z$\sim1.5$ (see \citealt{Mainieri2011,Bongiorno2014,Brusa2015}). 

The CO(2-1) transition in XID2028 is not detected down to a sensitivity of 0.23 mJy/beam over the full 770 km/s line width 
(corresponding to a 3$\sigma$ upper limit on the line integrated flux of 0.53 Jy km/s), from a separate, 3mm-band PdBI observation of XID2028 in October 2014 (M. Sargent, private communication). This suggests a near thermal CO-excitation state\footnote{r$_{31}\gsimeq1.0$ assuming CO[2-1] is thermalized, and r$_{31}\gsimeq0.9$ assuming CO[2-1] is sub-thermally excited $r_{21}=0.84$, standard value for MS objects}, and therefore r$_{31}$ around unity, i.e. larger than the standard value usually adopted for MS galaxies, more consistent with the QSO/Starburst scenario.
Given the complex nature of the system, we derive the CO(1-0) luminosity under the {\it conservative} assumption that r$_{31}$=0.7 (consistent with the constraints we have from millimeter data alone), and we will apply different $\alpha_{\rm CO}$ factors to derive molecular gas masses under the QSO/ULIRG and MS assumptions, discussing the implications of the findings in the different cases.
The inferred L’(CO[1-0]) luminosity for XID2028 (abbreviated as L'(CO) in the following) is therefore L’(CO)=2.6$\times10^{10}$ K km s$^{-1}$ pc$^2$ ($\sim1.2\times10^{6}$ L$_\odot$).

\subsection{Star Formation Efficiency}

\begin{figure*}[!t]
   \centering
 \includegraphics[width=8.7cm]{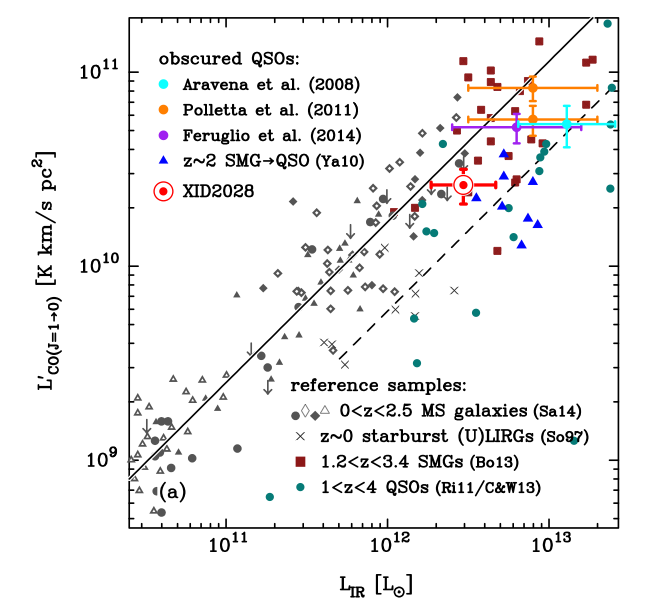}
 \includegraphics[width=8.0cm]{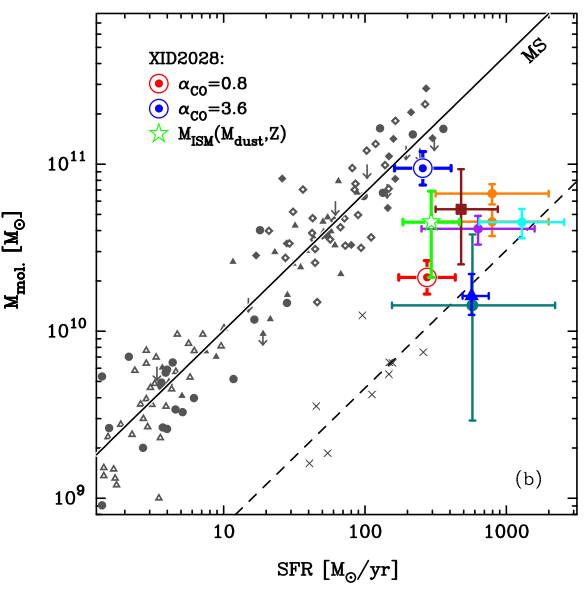}
     \caption{{\it Left panel}: L'(CO[1-0]) against the IR luminosity (8-1000$\mu$m) showing a compilation of MS galaxies at 0$<$z$<$2.5 from the \citet{Sargent2014} work (Sa14; grey symbols), local ULIRGs from from \citet[So97, crosses], SMGs from Bothwell et al. (2013; Bo13,  filled brown squares), 1$<$z$<$4 QSOs with data  from Riechers (2011) and Carilli \& Walter (2013; R11/C\&W13, light blue circles). 
The measurement for XID2028 obtained assuming $r_{31}=0.7$ is shown as a red circle. Obscured QSOs proposed to be in the transition phase presented in \citet{Aravena2008}, \citet{Polletta2011}, \citet{Feruglio2014} and the MIPS-selected sources from Yan et al. (2010) are also marked, as labeled. 
{\it Right panel}: Inverse, integrated Kennicutt-Schmidt relation between SFR and molecular gas mass. The color
points correspond to the position of XID2028 with different CO-to-H$_2$ conversion factors: $\alpha_{\rm CO}$=3.6 (top; blue),
and $\alpha_{\rm CO}$=0.8 (bottom; red). The green star shows the ISM mass inferred from the dust SED. All the values for XID2028 are slightly offset in the x-axis for clarity.  Other points are taken from the same samples presented in the left panel. For the Bo13, R11/C\&W13 and Ya10 samples we plot the median value with associated 16\% and 84\% percentiles. For the obscured QSOs all the authors used $\alpha_{\rm CO}$=0.8.
In both panels, the solid black line is the best-fit relation for MS galaxies, the dashed line defines the locus of strong SB galaxies with
approximately 15 times shorter depletion time (M$_{\rm mol}$/SFR) than MS galaxies \citep{Sargent2014}. 
              }
         \label{SFE}
   \end{figure*}

Figure~\ref{SFE} (left panel) shows  L'(CO) against the total  Infrared Luminosity (L$_{\rm IR}$, computed between 8-1000 $\mu$m) 
for XID2028 (red circle). The IR luminosity of XID2028 is very well constrained by Herschel/PACS+SPIRE data (logL$_{\rm IR}$=12.47; see \citealt{Brusa2015,Perna2015}) and has been estimated from fitting all bands with photometry with rest-frame wavelength $>50\mu$m  with \citet{Dale2002} Starbursts templates, using the same technique as in \citet{Santini2009}. 
Although recent works on FIR emission of AGN show that even the flux observed at rest frame  wavelengths longer than 60$\mu$m can be AGN-dominated (e.g. \citealt{Mullaney2011}), in XID2028 the QSO contribution is expected to be negligible, as shown in Figure 2 of \citet{Perna2015}, where the most recent SED fitting decomposition for this object is presented.
The observed IR luminosity corresponds to a SFR of $\sim270^{+50}_{-100}$ M$_\odot$ yr$^{-1}$, using the SFR-IR luminosity relation \citep{Kennicutt1998}, and taking into account the uncertainties on the flux normalization of the starburst component related to the AGN-host SED decomposition.

We compare this measurement with the compilation of low- and high-redshift normal star-forming  galaxies with measured $\alpha_{\rm CO}$ presented in \citet{Sargent2014} and the SMG-sample from \citet{Bothwell2013}, involving both outliers and galaxies consistent with the locus of the main sequence at their redshift. 
 We also plot the unobscured QSOs at 1$<z<$4 presented in \citet{Riechers2011}\footnote{In the case of lensed quasars, the values are corrected for the amplification, as reported in Riechers 2011.}. For these sources, the IR luminosities are extracted from the \citet{CW2013} compilation. 
Finally, in Figure 3 we show  SW022550 and SW022513 at z$\sim3.4$ \citep{Polletta2011}, ULASJ1539 at z$\sim2.5$  \citep{Feruglio2014}, 
and the MIPS-selected sources at z$\sim$2 from \citet{Yan2010}. All these systems have been proposed to be in the ``transition phase" between an heavily obscured Starburst phase and the unobscured QSO phase.

The SFE  of XID2028 (SFE$\sim110$) is on the lower side of the SFEs measured for high-z SMG and unobscured QSOs (SFE$\sim100-1000$). Instead, the SFE is consistent with  those reported (albeit with much larger uncertanties  due to the lack of a complete multiwavelength coverage and reliable measurements of L$_{\rm IR}$) in the obscured QSOs systems proposed to be in the ``transition phase" mentioned above.

%

\subsection{Molecular gas mass from CO data} 

Estimating the molecular gas mass based on the CO luminosity critically hinges on the CO-to-H2 conversion factor $\alpha_{\rm CO}$, defined as the ratio between the mass of molecular gas (M$_{\rm mol}$) to the integrated CO(1-0) luminosity ($\alpha_{\rm CO}$=M$_{\rm mol}$/L'(CO)[1-0] in units of M$_\odot$/(K km s$^{-1}$ pc$^{2}$)). This value depends on the ISM conditions, and two distinct assumptions are often adopted: 
$\alpha_{\rm CO}\sim4$ for extended SF disks/MS galaxies of solar metallicity, and $\alpha_{\rm CO}\sim0.8$ for compact  luminous systems (\citealt{Downes1998}; see \citealt{CW2013} and \citealt{Bolatto2013} for  in-depth discussions).

From a morphological point of view we do not have a clear classification on the properties of the host galaxy. Given that the HST image suffers from substantial extinction (A$_V\sim3$; see discussion in \citealt{Perna2015}) and, in any case, is dominated by the central active nucleus, it cannot be used for a reliable morphological analysis. However, no clear signatures of merging structures are visible in the rest-frame U-band. The low-resolution (with respect of HST) K-band image is instead consistent with both an elliptical galaxy and a spiral galaxy, possibly interacting with a north-east system (see Figure 2). 
Even if the MS is mainly populated by "normal" spiral and disk galaxies (see e.g. \citealt{Wuyts2011a}), we note that  XID2028 would lie among the population which occupies the upper envelope of the MS  at z$\sim1.5$. These galaxies may have also cushier light profiles, intermediate between disky galaxies and red and dead systems (see \citealt{Wuyts2011b}, their Figure 1, right panel).   In any case, if after point-source subtraction this galaxy were to be shown to have an early-type or disturbed host galaxy morphology, it would actually be highly consistent with the statistical findings of \citet{Wuyts2011b}.

In the vast majority of studies targeting SMGs, QSOs and ULIRGs systems (see e.g. \citealt{Aravena2008,Riechers2011_QSOz3,Polletta2011,Feruglio2014}, among others),  $\alpha_{\rm CO}=0.8$ has been adopted even in absence of  better information on the physical properties of the system (e.g. compactness of the source).  Under the assumption of  starbursts/QSO scenario,  we obtain for XID2028 a gas mass M$_{\rm mol}\sim$2.1$\pm0.4\times10^{10}$ M$_\odot$. 

To infer the molecular gas mass under the MS hypothesis, we consider a metallicity dependent conversion factor $\alpha_{\rm CO}$ (see e.g.\citealt{Genzel2012,Bolatto2013}). 
In the following we will assume for XID2028 a value of $12+log(O/H)=9.07$, the metallicity inferred from the so-called  Fundamental Metallicity Relation (FMR, \citealt{Mannucci2010}), that relates the metal content with the stellar mass and the SF of the galaxy independently of redshift \citep{Cresci2012}. 
Applying the relations  describing redshift-dependent variations of $\alpha_{\rm CO}$ in the SFR-M$_\star$ plane of \citet{Sargent2014} to XID2028 one would expect\footnote{A virtually identical conversion factor would be inferred using the relation between metallicity and $\alpha_{\rm CO}$ calibrated in \citet{Genzel2012} once the offsets between different metallicity calibrations are taken into account.} $\alpha_{\rm CO}\sim3.6$,
and the corresponding molecular gas mass would then be M$_{\rm mol}\sim9.5\pm1.9\times10^{10}$ M$_\odot$. 

The two values for M$_{\rm gas}$ inferred under the two different assumptions are plotted in  Figure 3b (with the  statistical errors associated to the line detection), where the molecular gas mass is shown as a function of the SFR (for the same samples presented in  Fig. 3a).

\subsection{Molecular gas mass from FIR emission} 

We  also adopt an independent method to compute the total gas mass in this source, using the dust mass derived  from FIR photometry.  For this purpose, we assume a metallicity-dependent
gas to dust ratio, following the calibration presented by \citet{Santini2014} and recently extended to AGN samples in the work by \citet{Vito2014_gas}. 
This estimate is independent from $\alpha_{\rm CO}$, although it depends on the metallicity ($Z$) of the system and on the 
assumptions that the dust-to-gas ratio scales linearly with $Z$ through a constant factor \citep{Draine2007}.

The dust mass is obtained via  SED decomposition of the AGN and host galaxy contributions,  using a combination of the \citet{Silva2004} AGN templates and the \citet{DraineLi2007} dust templates to fit the 100-500$\mu$m range. For objects at z$>$1, submillimeter data are in principle required to properly sample the dust emission free from AGN contamination. However, we note that for XID2028 the best fit SED decomposition performed following \citet{Vito2014_gas} is consistent with the upper limit of the continuum at 2mm (see Section 2.2).

We obtain a total dust mass of M$_{\rm dust}=7.7\pm4.2\times10^{8}$ M$_\odot$. 
Assuming the FMR metallicity (see above) and the \citet{Santini2014} calibration, this translates into a M$_{\rm gas}\sim4.5\pm2.4\times10^{10}$ M$_\odot$ (without considering the uncertainty on the dust-to-gas-ratio calibrations, e.g. a factor of $\sim2$, see \citealt{Sandstrom2013}). 
We note that using the \citet{Leroy2011} metallicity dependence of the dust-to-gas ratio would yield consistent results, within the errors. 

The value inferred from the dust fit approach is plotted as a green star in Fig. 3b.
If we use this estimate for M$_{\rm gas}$, and the observed L'(CO), we can derive an {\it effective} $\alpha_{\rm CO}$ for this source, $\alpha_{\rm CO(dust)}\sim2.4\times r_{31}$ ($\alpha_{\rm CO(dust)}\sim$1.7 given our adopted excitation correction r$_{31}$=0.7).

   \begin{figure*}[!t]
   \centering
 \includegraphics[width=18cm]{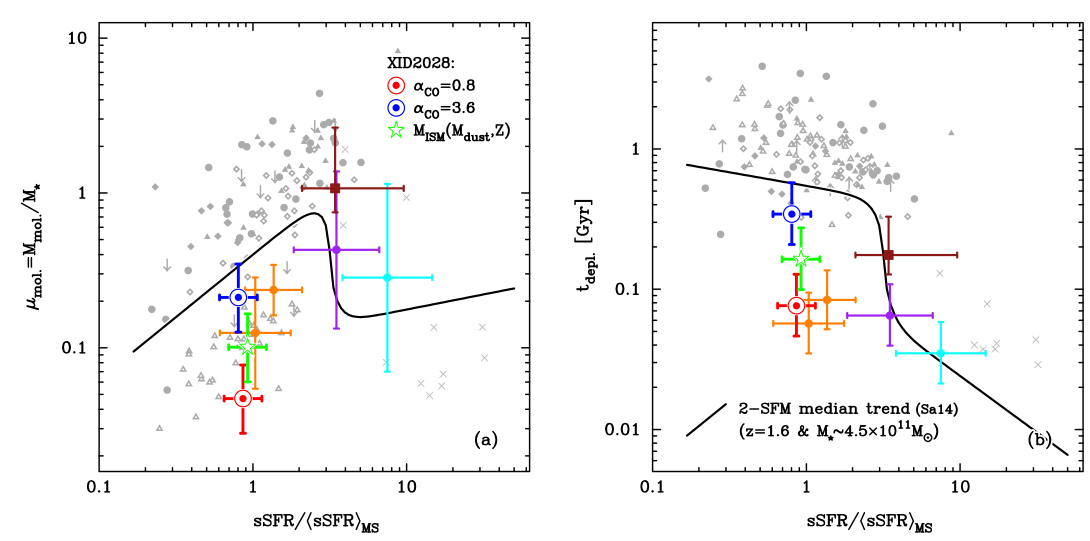}
      \caption{
Gas fraction $\mu_{\rm mol}$  {\it (Left panel)} and  depletion timescale  {\it (Right panel)} plotted versus the sSFR-excess for  the same samples and with the same color code presented in Figure 3. The values for XID2028 are slightly offset in the x-axis for clarity.  All quantities are normalized to the expected values for normal and starburst galaxies predicted by the calibration presented in \citet{Sargent2014}.  The black line traces the expected variation (median) with sSFR for a MS spiral galaxy with identical mass and redshift as XID2028 (see \citealt{Sargent2014}). The step at sSFR/$<$sSFR$>\sim$4 reflects the transition from the main sequence locus to the sSFR-regime where high-SFE starbursts dominate. XID2028  lies a factor $\sim$2 to $\sim$10 below the black line, i.e. it shows significant lower gas fraction and depletion time scale than those expected for the properties of the its host galaxy.
              }
         \label{fgas}
   \end{figure*}
%

\subsection{Gas fraction and depletion timescale}
The uncertainty in the derived gas mass from the CO data is dominated by the assumption in $\alpha_{\rm CO}$   (a factor of 4.5)  with respect to the statistical uncertainties (20\%). Given that the value derived from the dust fit is in between those for the two different assumptions in $\alpha_{\rm CO}$,  in the following we will refer to this value as our best estimate for the molecular gas mass, and those from the CO data under the MS and QSO/Starburst assumptions as upper and lower limit, respectively, i.e. M$_{\rm gas}=4.5 (1.7-11.4)\times10^{10}$ M$_\odot$, where the lower and upper limits in parenthesis take also into account the statistical uncertainties of the detection, and overall also the uncertainty in the assumed dust-to-gas ratio.
We note that the total gas mass inferred from the dust continuum fit includes both the molecular and atomic components. However, the atomic mass usually constitutes a negligible fraction of the total gas mass. 

The stellar mass of XID2028  is M$_{\star}\sim4.5\times10^{11}$ M$_\odot$ from the most recent SED fitting decomposition \citep{Perna2015}.  This value is a result of the inclusion in the multicomponent SED fitting of a mildly obscured QSO component, given that we observe the Broad Line Region (BLR) emission in the H$\alpha$ line complex \citep{Bongiorno2014}. 
In Section 2.2 we reported a dynamical mass M$_{\rm dyn}$$\sim$6$\pm2\times10^{11}$ M$_{\odot}$. Although the estimate of the dynamical mass suffer from large uncertainties, it is quite reassuring that it is consistent with the value we obtain from the sum of the stellar and molecular mass components (M$_{\rm tot}$$\sim4.6-5.6\times10^{11}$ M$_\odot$, taking into account the range of M$_{gas}$).

We can then calculate the molecular gas fraction, $\mu_{\rm mol}$, defined as the ratio of the molecular gas mass and the stellar mass ($\mu_{\rm mol}$=M$_{\rm mol}$/M$_{\star}$; see e.g. \citealt{Sargent2014,Genzel2014_gas}).  
Given the molecular gas masses inferred in the previous Section, the gas fraction translates into $\sim5$\%  for the QSO/Starburst and $\sim21$\% for the MS scenarios. The value from the dust mass measurement is in between these two estimates ($\sim10\%$). 
Similarly, we can estimate the depletion time scale (defined as M$_{\rm gas}$/SFR; e.g. the rate at which the gas is converted into stars) and we infer t$_{\rm depl}$=75, 340 and 160 Myr using the QSO/Starburst, MS and dust-fit derived gas masses, respectively. 

\subsection{Evidence for QSO feedback}

Figure~\ref{fgas} (left panel) shows the gas fraction in XID2028 for the three assumptions described above, plotted against 
the sSFR-excess with respect to the main sequence, e.g. sSFR/sSFR$_{\rm MS}$=$0.86$, where the mass- and redshift-dependence of the characteristic sSFR of MS galaxies follows the calibration in \citep{Sargent2014} which is based on a large compilation of literature data. 
In this plot we show the same samples used in Figure 3 (with the exception of unobscured QSOs and the MIPS selected sources 
for which no stellar mass estimates are available) and  we plot as a solid line the median trend 
with normalized sSFR, expected for a galaxy of the same mass and redshift of XID2028 
 (taken from the 2-Star Formation Mode description of normal and starbursting (off-MS) galaxies in \citealt{Sargent2014}).

Taking the best M$_{\rm gas}$ estimate  for our target, and even taking into account the uncertainty on $\alpha_{\rm CO}$ assumption,  XID2028  is among the objects with the lowest gas fraction for its sSFR detected so far in the high-z Universe and associated to normal star-forming galaxies (green star in Figure 4),
especially when compared to systems with similar masses (solid line).   
The $\mu_{\rm mol}$ is instead more similar to that expected for `Starburst' galaxies
of a similar mass and redshift (see value of black trend line at sSFR/<sSFR>$_{\rm MS}\gsimeq4$), 
but XID2028 does not share with these sources the same burst of star formation.
Instead, the gas fraction of XID2028 is similar to normal galaxies in the local universe (open triangles), despite its higher redshift.

An alternative way of visualizing the gas content and consumption is illustrated in 
the right panel of Figure~\ref{fgas}, where the depletion time scale
is plotted against the MS-normalised sSFR of the host galaxy.
Assuming our best M$_{\rm gas}$ estimate, 
XID2028 lies at shorter depletion time scales with respect to MS galaxies (at any redshift),
i.e. it is consuming its residual gas 
more rapidly than normal star-forming galaxies.
This qualifies XID2028 as a clear outlier with respect to the average population, and a rare object, consistent with the hypothesis that it is caught in the very short transition phase in which the QSO feedback is released. 

Similar conclusions can be reached examining the position of our source with respect to the Kennicutt-Schmidt relation \citep{Kennicutt1998}: assuming the physical scales inferred in Section 2.2, and that the molecular gas and the SF episodes are distributed uniformly over this region, XID2028 would lie slightly above (a factor $\sim2.5$) the correlation observed for normal and starburst galaxies. However, the SFR density measured in this way is to be considered a lower limit, given that the SF regions seem to be patchy (see Cresci et al. 2015). Therefore, XID2028 would further deviate above the K-S relation, towards regions of short depletion timescales. 

It is important to note that, even when using the MS assumption, the molecular gas fraction and depletion timescale would be considerably lower than  those expected for systems of the same host galaxies properties of XID2028 (blue point in Figure 4). In particular, the depletion timescale observed in XID2028 for the MS scenario is a factor of $\sim2$ lower than the expectations of \citet{Sargent2014} and a factor $\sim3$ lower than that obtained by the parameterization of MS and off-MS galaxies presented in \citet[using their {\it global fit} we expect for XID2028 $\rm{t_{dep}(G15)_{global}}\sim970$ Myr]{Genzel2014_gas}. 
The discrepancy with the calibrations is more extreme if the values obtained in the QSO scenario are adopted (red circles in Figure 4). 
We also note that our chain of assumptions in deriving M$_{\rm gas}$ has been very conservative. For example, we used r$_{31}$=0.7 instead of $r_{31} \gsimeq 0.9$ as suggested by the non detection of CO[2-1] emission, which would have instead provided a 20\% smaller CO[1-0] flux. This conservative assumption also compensates a possible overestimate of the value of the CO[3-2] flux, which could result from measuring the line flux at the phase centre rather than at the slightly offset centroid. The result of the lack of molecular gas in XID2028, with M$_{\rm gas}\lsimeq10^{11}$ M$_\odot$, is therefore quite robust. 

A short depletion time scale with respect to MS galaxies has also been found for SMGs in the \citet{Bothwell2013} sample, and  other AGN/Starburst systems plotted in Figure 4 \citep{Aravena2008,Polletta2011,Feruglio2014}.  \citet{Yan2010} also reported a short depletion timescale of $\sim$40 Myr for the sample of MIPS-selected ULIRGs.
The short depletion time scale in SMGs has been interpreted as higher star formation efficiency in the galaxy (e.g. \citealt{Genzel2010,Daddi2010}), probably due to higher density of the ISM in these compact systems. These may also be the case for  ULASJ1534 and the COSBO11, which have sSFRs comparable to SMGs, and for which we expect compact gas reservoirs. 

Instead, in XID2028 a significant fraction of the gas is expected to be already expelled from the galaxy. The SF is then probably maintained only in the denser environments, less affected by the negative feedback, and possibly enhanced by positive feedback due to the outflow induced pressure (e.g. \citealt{Silk2013}).
The fact that XID2028 has a smaller gas reservoir and shorter depletion time than that measured for MS galaxies of similar sSFR therefore constitutes a new probe, 
in  addition to the analysis presented in \citet{Cresci2015} based on NIR data, that QSO feedback in the form of powerful outflows is able to affect star formation in the host and expel a significant fraction of gas from the host galaxy.

\section{Summary}
We presented the first molecular line luminosity measurement, via \co observations obtained at the PdBI interferometer, in a luminous obscured QSO at z$\sim$1.5. The target is thought to be in the `blow-out' phase, and the presence of a powerful outflow with significant impact on the host galaxy has been unveiled through previous NIR observations \citep{Perna2015,Cresci2015}. We complemented the PdBI data with FIR dust fitting, and report the following results: 

\begin{itemize}

\item[$\bullet$]  We measure a SFE ($\simeq$110) at the lower end of those reported in the literature for a large number of QSOs and Starburts/SMG galaxies (see \citealt{Iono2009,CW2013}), and consistent with that inferred for obscured QSOs at higher redshift; 

\item[$\bullet$]
We infer a molecular gas mass (M$_{\rm mol}$) in the range $2.1\pm0.4\div9.5\pm1.9\times10^{10}$ M$_\odot$ 
applying  the QSO/Starburst or MS conversion factors to the measured L'CO line luminosity, respectively, and a total gas mass M$_{\rm gas}\sim4.5\times10^{10}$ M$_\odot$ from dust continuum fitting; 

\item[$\bullet$]
A value for the molecular gas mass $<10^{11}$ M$_\odot$ is also remarkably consistent with our estimates of the dynamical mass 
through spectroastrometric methods (see Section 2.2), given the high stellar mass of XID2028;

\item[$\bullet$]
We also infer a molecular gas fraction  $\mu_{\rm mol}\sim5-20$\%. 
This translates into a gas depletion time scale t$_{\rm depl}\sim$70-340 Myr, depending on the assumptions on $\alpha_{\rm CO}$ (see Figure 4). 

\item[$\bullet$]
The value of t$_{\rm depl}$ is considerably lower ($\lsimeq30$\%) than those observed in systems hosted in similar massive (M$_{\star}>10^{11}$ M$_\odot$) MS galaxies (MS-normalised sSFR$\sim1$), and consistent with  those observed for SMGs and for the other few systems proposed to be in the transition phase.

\end{itemize}

\par\noindent

We propose that in XID2028 the QSO wind, detected in the ionised gas component out to 10-kpc scales, has already removed most of the molecular gas from the host galaxy. 
All the observational constraints  (low molecular gas content, lowest $\mu_{\rm mol}$ at a fixed sSFR when compared to M$_\star>10^{11}$ M$_\odot$ systems, and lowest sSFR at a fixed $\mu_{\rm mol}$) are consistent with such a scenario,  where the gas in the host galaxy of XID2028 
is indeed already depleted/dispersed by the effects of the strong QSO feedback (see also \citealt{Coppin2008} and \citealt{Yan2010} for similar interpretation). 
In dense regions (e.g. clumpy M$_{gas}$ reservoirs), possibly located at the edge of the outflow cavity \citep{Cresci2015}, the residual gas is converted into stars at a high rate  similar to that observed in SMGs,  where the low depletion time scale is indeed ascribed to the efficient SF triggered in dense and compact gas reservoirs. 

The measure of the intensity of the \co emission in XID2028 represents a first step towards a  mapping experiment using high spatial resolution  to study the morphology and the kinematics of the molecular gas reservoir and 
of the clumpy structures in the distribution of SF regions seen in HST and SINFONI maps. 
Sensitive ALMA and/or NOEMA observations of XID2028 will finally 
give the spatial resolution to locate molecular clouds  (see, e.g., \citealt{Aravena2014}) and reveal any possible molecular outflow component.

\begin{acknowledgements}
Based on observations carried out under project number X--8 with the IRAM PdBI 
Interferometer. IRAM is supported by INSU/CNRS (France), MPG (Germany) and IGN
(Spain). We gratefully acknowledge the allocation of IRAM DDT time, and we thank the staff of the IRAM Observatory for their support of this program.
MB, MP and GL acknowledge support from the FP7 Career Integration Grant ``eEASy'' (``SMBH evolution through cosmic time: from current surveys to eROSITA-Euclid AGN Synergies", CIG 321913). 
MB gratefully acknowledges fundings from the DFG cluster of excellence `Origin and Structure of the Universe' (www.universe-cluster.de).
We acknowledge financial support from INAF under the contracts PRIN-INAF-2011 (``Black Hole growth and AGN feedback through cosmic time"), PRIN-INAF-2012 (``The Lifecycle of early Black Holes'') and PRIN MIUR 2010-2011 (``The dark Universe and the cosmic evolution of baryons"). 
We thank Dennis Downes and Andrea Comastri for enlightening discussion. We thank the anonymous referee for his/her interest towards the results of our work, a very careful reading of the paper and useful suggestions which improved the presentation of the results. 

\end{acknowledgements}


\end{document}